\documentclass[10pt,letterpaper,twocolumn]{article} 
\usepackage{ol2} 
\usepackage{hyperref,cite}

\usepackage{graphicx,ulem}
\usepackage{amsmath,amstext,amssymb}

\newcommand{\E}{\mathbf{E}}

\newcommand{\Pol}{\mathbf{P}}

\newcommand{\evac}{\varepsilon_0}
\newcommand{\chit}{\chi^{(2)}}

\newcommand{\kz}{k}
\newcommand{\bp}{\kz^{(1)}}
\newcommand{\bpp}{\kz^{(2)}}

\newcommand{\ld}{L_{\rm D,1}}
\newcommand{\lshgb}{L_{\rm SHG}}
\newcommand{\lmax}{z_{\rm opt}}
\newcommand{\lkerrb}{L_{\rm Kerr}}

\newcommand{\tin}{T_{1,\rm in}}

\newcommand{\Iin}{I_{1,\rm in}}
\newcommand{\Ein}{{\mathcal E}_{1,\rm in}}

\newcommand{\neff}[1]{n_{#1}}%
\newcommand{\deff}{d_{\rm eff}}%
\newcommand{\nshgb}{N_{\rm SHG}}%
\newcommand{\neffs}{N_{\rm eff}}%
\newcommand{\neffc}{N_{\rm eff,c}}%
\newcommand{\nkerrb}{N_{\rm Kerr}}%
\newcommand{\gammab}{\gamma_1^I}
\newcommand{\Lm}{\hat{\mathcal{L}}}
\newcommand{\Dm}{\hat{\mathcal{D}}}
\newcommand{\Rm}{\mathcal{R}}

\newcommand{\lgvm}{L_{\rm GVM}}
\renewcommand{\onlinecite}{\cite} 
\begin{document}
\twocolumn[ 
\title{Scaling laws for soliton pulse compression by cascaded
  quadratic nonlinearities}

\author{M. Bache$^{1,*}$, J. Moses$^{2,3}$, and F. W. Wise$^2$} 
\address{$^1$COM$\bullet$DTU, 
Technical University of Denmark,  Bld. 345v, DK-2800 Lyngby, Denmark.
\\
$^2$Department of Applied and Engineering Physics, Cornell
  University, Ithaca, New York 14853.
\\$^3$Currently with: Research Laboratory of Electronics,
  Massachusetts Institute of Technology, Cambridge, MA 02139
\\$^*$Corresponding author: bache@com.dtu.dk}

 \date{\today}
\begin{abstract}
  We present a detailed study of soliton compression of ultra-short
  pulses based on phase-mismatched second-harmonic
  generation (\textit{i.e.}, the cascaded quadratic nonlinearity) in
  bulk quadratic nonlinear media. The single-cycle propagation
  equations in the temporal domain including higher-order nonlinear
  terms are presented. The balance between the quadratic (SHG) and the
  cubic (Kerr) nonlinearity plays a crucial role: we define an
  effective soliton number -- related to the difference between the
  SHG and the Kerr soliton numbers -- and show that it has to be
  larger than unity for successful pulse compression to take place.
  This requires that the phase mismatch be below a critical
  level, which is high in a material where the quadratic nonlinearity
  dominates over the cubic Kerr nonlinearity.  Through extensive
  numerical simulations we find dimensionless scaling laws, expressed
  through the effective soliton number, which control the behaviour of
  the compressed pulses. These laws hold in the stationary regime, in
  which group-velocity mismatch effects are small, and they are
  similar to the ones observed for fiber soliton compressors.  The
  numerical simulations indicate that clean compressed pulses below
  two optical cycles can be achieved in a $\beta$-barium borate
  crystal at appropriate wavelengths, even for picosecond input
  pulses.
\end{abstract}

\ocis{320.5520, 320.7110, 190.5530, 190.2620, 190.4400}
\maketitle
]
\section{Introduction}
\label{sec:Introduction}

Second-harmonic generation (SHG) in the limit of large phase
mismatch gives rise to the so-called cascaded
$\chit:\chit$-nonlinearity, in which weak conversion to the second
harmonic (SH) occurs, while a Kerr-like nonlinear phase shift is
induced on the fundamental wave (FW)
\cite{desalvo:1992,stegeman:1993}.  An elegant theoretical view of
this process is that in the cascading limit the system can be
described by a nonlinear Schr{\"o}dinger equation (NLSE) for the FW
alone \cite{menyuk:1994}. The induced FW nonlinear phase shift then
comes from an effective self-phase modulation arising from the phase
mismatch. The phase shift can be quite large and negative, since the
phase mismatch determines the sign and magnitude of the effective
cubic nonlinearity.  Such a \textit{self-defocusing} nonlinearity can
be used to compress a pulse when combined with normal dispersion
\cite{liu:1999,ashihara:2002,ashihara:2004,zeng:2006,moses:2006,xie:2007},
and problems normally encountered due to \textit{self-focusing} in
cubic media are avoided.  Thus, having no power limit, a
self-defocusing compressor can create high-energy near single-cycle fs
pulses in bulk media \cite{liu:1999,moses:2006}. The exploited
compression mechanism is the periodic behaviour of higher-order
temporal solitons of the NLSE.  They oscillate in temporal duration
upon propagation, and the optimal compressor length is when the pulse
width is the narrowest.

As the compressor scheme exploits an effective self-defocusing cubic
term from cascaded quadratic effects, the compressor will naturally be
affected by the self-focusing cubic nonlinearity inherent to any
transparent material. This detrimental cubic nonlinearity must be
counterbalanced and then exceeded to achieve compression
\cite{ashihara:2002,moses:2006}. In this work we systematically
demonstrate theoretically and numerically how this can be expressed
conveniently using the soliton number formalism known from the NLSE in
fiber optics \cite{agrawal:1989,agrawal:2001}.  Firstly, we show that
the SHG soliton number \cite{moses:2006} $\nshgb$ must outbalance the
Kerr soliton number $\nkerrb$ before compression takes place. More
precisely, the \textit{effective} soliton number
$\neffs=\sqrt{\nshgb^2-\nkerrb^2}$ must be larger than unity.  This
can be achieved by adjusting the phase mismatch, but only if the
quadratic material nonlinearity is sufficiently stronger than the
cubic Kerr nonlinearity. Secondly, when compression is successful, the
compression factor, pulse quality, and optimal compressor length
follow certain scaling laws. Such empirical scaling laws expressed
through $\nkerrb$ were previously given for the NLSE
\cite{mollenauer:1983,tomlinson:1984,dianov:1986}. Here we present
detailed numerical simulations of pulse propagation in a commonly-used
nonlinear material, $\beta$-barium borate (BBO). We find
general dimensionless scaling laws quite similar to the ones from the
NLSE, except that one must express them through $\neffs$.  The
simulations span a very wide input-pulse parameter space and the
scaling laws therefore provide an important tool for experimental
situations.

A major obstacle to the cascaded-quadratic soliton compressor is the
group-velocity mismatch (GVM) between the FW and SH pulses. GVM
induces a Raman-like perturbation that distorts the compression,
resulting in asymmetric pulses and pulse splitting
\cite{liu:1999,ilday:2004,moses:2006}.  Two distinct regimes exist: in
the \textit{stationary} regime the GVM-induced Raman-like
perturbations are weak, resulting in clean compressed pulses. In the
\textit{nonstationary} regime, the GVM-induced perturbations are
strong, and poor pulse compression is observed. Moreover, the
compression behavior deviates from the NLSE scaling laws.  Here we
focus on the stationary regime, where cleanly compressed few-cycle
pulses may be obtained, and summarize the guidelines for avoiding
nonstationary behavior. Subsequent publications will focus on
compression in the nonstationary regime in greater detail.  In
particular, we have recently devised a nonlocal theory that
quantitatively defines the boundary between the stationary and
nonstationary regimes \cite{bache:2007a}.

Finally, when describing (ultra)-short pulse propagation in nonlinear
media it is generally important to describe both temporal and spatial
effects as they will inherently be interlinked. Examples include
space-time focusing in bulk media \cite{brabec:1997}, and strong
modification of the dispersion and nonlinear mode overlap in waveguide
geometries.  However, the simplified 1D temporal description presented
here is often an adequate starting point for understanding the general
temporal behaviour of the system during compression.  In this work we
consider interaction in a bulk material (no wave guiding) and we
neglect diffraction (\textit{i.e.}, the beam must not be focused too
tightly). In subsequent publications we will address the behaviour in
a wave-guiding geometry such as a fiber, as well as in bulk geometry
including the spatial effects.

\section{Generalized propagation equations for ultra-short pulses}
\label{sec:Gener-prop-equat}

Below we present the generalized bulk SHG propagation equations,
including both quadratic and cubic nonlinear effects.  We aim to
investigate compressed pulses that may be near-single optical cycle in
duration, so we must use the propagation equations derived without
approximations that directly impart constraints on the pulse duration
and bandwidth. Such SHG equations have been derived elsewhere
\cite{moses:2006b} using the slowly evolving wave approximation (SEWA)
\cite{brabec:1997}. Here we recast these equations in the framework of
cascaded quadratic soliton compression as to define all critical
experimental parameters. We also include generalized cubic
nonlinearities from interaction of two fields with different
frequencies (the FW and SH), see App.~\ref{sec:cubic-nonl-resp}.

The investigation is a temporal study only.  We neglect transverse
spatial effects, assuming that diffraction plays a negligible role on
the length scale of the compressor, \textit{i.e.}, that each point on
the beam may be considered a plane wave.  Previously-reported
quadratic soliton compressors have lengths on the order of a few
centimeters
\cite{liu:1999,ashihara:2002,ashihara:2004,zeng:2006,moses:2006}.  For
a Gaussian beam at near-infrared wavelengths, this implies that our
assumptions are valid for beam waists larger than approximately 100
$\mu$m.  Section~\ref{sec:Other spatial geometries} will discuss
considerations of compression with transverse-spatially varying
profiles. Finally note that in our notation, a primed variable
is always dimensionless.

\subsection{The dimensional form}
\label{sec:dimensional-form}

In SHG two FW photons of frequency $\omega_1$ combine to give a photon
of frequency $\omega_2=2\omega_1$ (the SH). We consider scalar fields
and either a Type-0 or Type-I SHG geometry, in which the two FW
photons have identical polarization \cite{dmitriev:1999}. In a bulk
medium, the generalized SEWA propagation equations for the coupled
electric fields are
\begin{subequations}
\label{eq:shg-bulk-E}
\begin{align}
  \label{eq:shg-bulk-fh-raman}
  &\Lm_1E_1+\kappa_{\rm SHG,1}^E\hat S_1E_1^*E_2e^{i\Delta k
    z}+\kappa_{\rm Kerr,1}^E \times
 \\  &\left[(1-f_R)\hat
  S_1E_1\left(|E_1|^2+B|E_2|^2\right)+f_R\Rm_1(\tau)\right]=0,
\nonumber\\
  \label{eq:shg-bulk-sh-raman}
  &\Lm_2E_2+\kappa_{\rm SHG,2}^E\hat S_2E_1^2e^{-i\Delta k
  z} +\kappa_{\rm Kerr,2}^E\times
\\  &\left[(1-f_R)\hat 
  S_2E_2\left(|E_2|^2+B|E_1|^2\right)+f_R\Rm_2(\tau)\right]=0,
\nonumber
\end{align}
\end{subequations}
where $E_j=E_j(z,t)$.  For a Type-0 geometry the cross-phase
modulation (XPM) coefficient is $B=2$, while for a Type-I geometry
$B=2/3$ because the FW and SH have orthogonal polarization
\cite{agrawal:2001}. We assume that the FW and SH spectra do not
significantly overlap, an assumption generally valid down to
single-cycle pulse durations. The linear propagation operators are
\begin{subequations}
  \label{eq:Lprop}
\begin{align}
  \label{eq:Lprop_fh}
  \Lm_1&\equiv i\frac{\partial}{\partial z}
+\Dm_1, 
\\
\label{eq:Lprop_sh}
  \Lm_2&\equiv i\frac{\partial}{\partial
    z}-id_{12}\frac{\partial} {\partial \tau} +\Dm_{2,\rm eff},
\end{align}
\end{subequations}
where $\hat D_j$ are the dispersion operators
\begin{align}
\label{eq:Dprop}
\Dm_j\equiv\sum_{m=2}^{\infty}i^m \frac{\kz_j^{(m)}}{m!}\frac{\partial^m}{\partial
  \tau^m} ,
\end{align}
Here $k_j=n_j\omega_j/c$, where $n_j$ is the refractive index. The
fields are in the frame of reference traveling with the FW group
velocity $v_{g,1}$ by the transformation $\tau=t-z/v_{g,1}$. This is
the origin of the GVM-term $d_{12}=1/v_{g,1}-1/v_{g,2}$, where
$v_{g,j}^{-1}=\bp_j$.  $\kz_j^{(m)}\equiv \partial^m \kz_j/\partial
\omega^m|_{\omega=\omega_j}$ accounts for dispersion, and $\Delta
k\equiv k_2-2k_1$ is the phase mismatch.  Note the unfamiliar term in
Eq.~(\ref{eq:Lprop_sh}), which is an effective SH dispersion operator
$\Dm_{2,\rm eff}= \Dm_2+ \hat S_2^{-1}\frac{d_{12}^2}{2k_2}
\frac{\partial^2 }{\partial\tau^2}$ [see
Eq.~(\ref{eq:D2-eff-dimless})], which deviates from $\Dm_2$ due to GVM
and self-steepening in the SEWA model \cite{moses:2006b},
see App.~\ref{sec:Appendix}.

The quadratic nonlinear coefficient is $\kappa_{{\rm SHG},j}^E\equiv
  \chit\omega_1/2c\neff{j}
  =\deff\omega_1/c\neff{j}$
where 
$\chit$ is the value of the
quadratic nonlinear tensor along the polarization direction of the
interacting waves, and as is typical we take $\deff\equiv \chit/2$.
Dispersion of $\chit$ is assumed negligible.

The cubic nonlinear coefficient is $ \kappa_{{\rm
    Kerr},j}^E=\omega_jn_{{\rm Kerr},j}/c $. 
The Kerr nonlinear refractive index is $n_{{\rm Kerr},j}\equiv
3{\rm Re}(\chi^{(3)})/8 \neff{j}$, where  
$\chi^{(3)}=\chi^{(3)}_{xxxx}$, see App.~\ref{sec:cubic-nonl-resp}.
We neglect two-photon absorption, implying
${\rm Im}(\chi^{(3)})=0$, which holds as long as
the SH frequency spectrum lies below the two-photon absorption edge of
the medium. Note, the usual notation for $n_{{\rm Kerr},j}$ is 
$n_2$, but we reserve the subscript 2 for the SH.

Self-steepening effects are modelled by the operator $\hat S_j\equiv 1+\frac{i}{\omega_j}\frac{\partial}
  {\partial \tau}$,
It approaches unity when bandwidths $\Delta \omega_j$
are small compared to the carrier frequencies $\omega_j$, \textit{i.e.}, for
pulses longer than roughly 10 optical cycles.

Finally, $\Rm_j(\tau)$ describes the vibrational Raman
response of the cubic nonlinearity, see
App.~\ref{sec:cubic-nonl-resp}. We henceforth set
$f_R=0$, because Raman scattering plays a negligible role under the
relevant experimental conditions.  

We now scale the fields by defining $A_j\equiv E_j\sqrt{\varepsilon_0
  n_j c/2}$ so $|A_j|^2$ gives the beam intensity $I_j$.
Eqs.~(\ref{eq:shg-bulk-E}) become
\begin{subequations}\label{eq:shg-bulk-int}
\begin{align}
  \label{eq:shg-bulk-fh-int}
\Lm_1A_1
  &+\kappa_{\rm SHG}^I\hat S_1A_1^*A_2e^{i\Delta k z}
\\\nonumber&+\gammab \hat S_1
  A_1\left(|A_1|^2+B\bar{n}|A_2|^2\right)=0,\\ 
  \label{eq:shg-bulk-sh-int}
\Lm_2A_2&+\kappa_{\rm SHG}^I\hat S_2A_1^2e^{-i\Delta k
  z}
\\\nonumber&+2\bar{n}^2\gammab\hat
  S_2A_2\left(|A_2|^2+B\bar{n}^{-1}|A_1|^2\right)=0.
\end{align}
\end{subequations}
The nonlinear coefficients due to the intensity scaling are $
\kappa_{\rm SHG}^I\equiv \sqrt{2\omega_1^2d_{\rm eff}^2/
  \neff{1}^2\neff{2}\varepsilon_0c^3}$, and $\gammab\equiv\omega_1
n_{{\rm Kerr},1}^{I}/c$.
Here the intensity-version of the Kerr nonlinear refractive
index with unit $[{\rm m^2/W}]$ is $ n_{{\rm
    Kerr},1}^{I}\equiv 2n_{{\rm Kerr},1}/\neff{1}\varepsilon_0 c
=3{\rm Re}(\chi^{(3)})/4 \neff{1}^2\varepsilon_0 c.$
Note the connection to the fiber NLSE
(where $A_j$ is scaled to the power);
$\gammab$ is equivalent to the
parameter \cite{agrawal:1989} $\gamma_1=\omega_1 n_{\rm Kerr,1}^I/c
A_{\rm eff}$, where $A_{\rm eff}$ is the effective mode area of the
FW. Finally, $\bar n\equiv \neff{1}/\neff{2}$ which is
typically close to unity, except when $\Delta k$ is large. 

\subsection{The dimensionless form}
\label{sec:Deriv-dimensionless}

We now rescale space and time as $z'\equiv
z/\ld$, and $\tau'=\tau/\tin$,
where $ \ld\equiv \tin^2/|\bpp_1|$ and $\tin$ are the FW dispersion
length
and input pulse duration. When compressing a FW
pulse, no SH is launched, so both fields are scaled 
to the peak FW input intensity $\Iin\equiv |A_1(0,0)|^2$,
\textit{i.e.}, $U_j=A_j/\sqrt{\Iin}$. Thus,
$U_1=E_1/\Ein$, where $\Ein\equiv
|E_1(0,0)|$, and Eqs.~(\ref{eq:shg-bulk-int}) become
\begin{subequations}\label{eq:shg-bulk-U}
  \begin{align}
\label{eq:shg-bulk-fh-U}
  \Lm_1'U_1
  &+\nshgb \sqrt{|\Delta k'|} \hat S_1'U_1^*U_2e^{i\Delta
  k'z'}
\\&+\nkerrb^2\hat S_1'U_1\left(|U_1|^2+B\bar{n}|U_2|^2\right)=0,
\nonumber\\
\label{eq:shg-bulk-sh-U}
  \Lm_2'U_2&+\nshgb \sqrt{|\Delta k'|} \hat S_2' U_1^2e^{-i\Delta k'z'} 
\\&
  +2\bar{n}^2\nkerrb^2
\hat S_2'U_2\left(|U_2|^2+B\bar{n}^{-1}|U_1|^2\right)    
=0,
\nonumber
  \end{align}
\end{subequations}
where $\nshgb$ and $\nkerrb$, the dimensionless quadratic and cubic
soliton numbers, respectively, are defined as  
\begin{subequations}
\begin{align}
  \label{eq:Nshg-bulk}
  \nshgb^2&=
\ld\frac{\omega_1^2\deff^2 
  \Ein^2}{c^2 \neff{1}\neff{2}|\Delta k|},
  \\
  \nkerrb^2&=
\ld\gammab\Iin=\ld\omega_1n_{{\rm Kerr},1}^{I}\Iin/c,
  \label{eq:Nkerr-bulk}
\end{align}
\end{subequations}
and the dimensionless phase mismatch is $\Delta k'\equiv \Delta k\ld$.
The parameter $\nshgb$ is deliberately chosen for a soliton
compression scenario, as will become clear later.

Equations~(\ref{eq:Lprop}) on dimensionless form are $  \Lm_1'\equiv i\frac{\partial}{\partial z'}
+\Dm'_1,$ and  $\Lm_2'\equiv i\frac{\partial}{\partial
    z'}-id_{12}'\frac{\partial} {\partial \tau'} +\Dm'_{2,\rm eff}$,
  and $\Dm'_{2,\rm eff}$ is given by Eq.~(\ref{eq:D2-eff-dimless-final-app}).
The dimensionless dispersion operators and corresponding coefficients are $\Dm_j'\equiv \sum_{m=2}^{m_d}i^m \delta_j^{(m)}\frac{\partial^m}{\partial
  \tau'^m}$, 
and $d_{12}'\equiv
d_{12}\tin/|\bpp_1|$, and $\delta_j^{(m)}\equiv \ld
\kz_j^{(m)}(\tin^{m-2}|\bpp_1|m!)^{-1} $
Finally, using $ s'\equiv(\omega_1\tin)^{-1}$ we have the steepening
operators with dimensionless time $\hat S_1'\equiv
1+is'\frac{\partial} {\partial \tau'}$, and $\hat S_2'\equiv
1+i\frac{s'}{2}\frac{\partial} {\partial \tau'}.$


\subsection{The soliton numbers}
\label{sec:soliton-numbers}

We now turn our attention to the soliton numbers as defined in
Sec.~\ref{sec:Deriv-dimensionless}. We will argue -- and later show
with numerical simulations -- that the effective soliton number
$\neffs=\sqrt{\nshgb^2-\nkerrb^2}$ can be used to describe the system
in general. Indeed, $\neffs$ characterizes the outcome for any given
input state of the system.

First, consider the NLSE in absence of quadratic nonlinearities. We
include for instruction the Raman terms and ignore the steepening
terms [Eq.~(\ref{eq:shg-bulk-Raman-app-fh}) with $\kappa_{{\rm
    SHG},1}^E = 0$ and $\hat S_j=1$]. In dimensionless form it reads:
\begin{multline}
  i\frac{\partial U_1}{\partial z'}-\frac{{\rm sgn}(\bpp_1)}{2}
  \frac{\partial^2 U_1}{\partial \tau'^2}
\\  +\nkerrb^2\left[|U_1|^2U_1-\tau'_{R}U_1
  \frac{\partial |U_1|^2}{\partial \tau'}\right] =0,  
  \label{eq:nlse-fh}
\end{multline}
which holds when the pulse duration is much slower than the Raman
response, and $\tau'_R=\tau_R/\tin$, where $\tau_R$ is a
characteristic Raman vibrational response time \cite{agrawal:1989},
see App.~\ref{sec:cubic-nonl-resp}. The Kerr soliton
number~(\ref{eq:Nkerr-bulk}) we express as $\nkerrb \equiv
\sqrt{\ld/\lkerrb}$,
where $\lkerrb = (\gammab\Iin)^{-1}$ is the characteristic
Kerr length.  Equation~(\ref{eq:nlse-fh}) governs soliton compression
in cubic nonlinear media, and the compression characteristics are well
understood to be critically dependent on $\nkerrb$
\cite{agrawal:1989,agrawal:2001,mollenauer:1983,tomlinson:1984,dianov:1986}.
We stress that soliton compression using a pure Kerr self-focusing
nonlinearity requires anomalous dispersion ${\rm sgn}(\bpp_1)<0$.

Likewise, the quadratic soliton number $\nshgb$ is a critical
parameter in cascaded quadratic soliton compression that as we shall
now see has a role analogous to the cubic soliton number of the NLSE.
Using perturbative methods one can reduce the
dimensionless SHG equations~(\ref{eq:shg-bulk-U}) -- in absence of
steepening and cubic nonlinear terms --
to a single equation for the FW analogous to Eq.~(\ref{eq:nlse-fh})
when the phase mismatch $\Delta k$ is suitably large
\cite{menyuk:1994,ilday:2004}.  The approximate equation for the FW
field can for soliton compression purposes be cast as
\cite{moses:2006}
\begin{multline}
  i\frac{\partial U_1}{\partial z'}-\frac{{\rm sgn}(\bpp_1)}{2}
  \frac{\partial^2 U_1}{\partial \tau'^2}
  -{\rm sgn}(\Delta k)\times
\\
  \label{eq:fh-shg-nlse-moses}
  \nshgb^2\left[|U_1|^2U_1+is_{12}\tau'_{R,{\rm SHG}}|U_1|^2
    \frac{\partial U_1}{\partial \tau'}\right] =0,
\end{multline}
where $s_{12}={\rm sgn}(d_{12})$. In
analogy to the cubic nonlinearity, a characteristic
length for the cascaded quadratic nonlinearity $\lshgb\equiv c^2 \neff{1}\neff{2}|\Delta k|/\omega_1^2\deff^2
\Ein^2$, 
such that $\nshgb \equiv
 \sqrt{\ld/\lshgb}$.
The similarity to Eq.~(\ref{eq:nlse-fh}) is clear, and explains the
choice of $\nshgb$ in Eq.~(\ref{eq:Nshg-bulk}). Similar to
$\nkerrb$ for the NLSE, $\nshgb$ is namely a critical parameter in cascaded
quadratic soliton compression.  Several other features should be noted.
First, the sign of the phase mismatch controls the sign of the induced
cubic nonlinearity.  For the purpose of soliton compression with
normal dispersion ${\rm sgn}(\bpp_1)>0$, we require $\Delta k>0$ in
order to have self-defocusing, or negative nonlinearity.  Second, the
Raman-like term in Eq.~(\ref{eq:fh-shg-nlse-moses}) is important in
the compression dynamics, as will be explained in
Sec.~\ref{sec:stationary-regime-1}.

Equations (\ref{eq:nlse-fh}) and (\ref{eq:fh-shg-nlse-moses}) consider
the cases where either the cubic or the quadratic nonlinearity is
significant. However, we must consider both orders of nonlinearity, so
using the same perturbation methods, we reduce
Eqs.~(\ref{eq:shg-bulk-U}) to an approximate equation for the FW.  To
first order, and neglecting steepening effects, we find
\begin{multline}
  \label{eq:fh-chi2-and-chi3}
  i\frac{\partial U_1}{\partial z'}-\frac{{\rm sgn}(\bpp_1)}{2}
  \frac{\partial^2 U_1}{\partial \tau'^2}
\\
  -{\rm sgn}(\Delta k)
\nshgb^2\left[|U_1|^2U_1+is_{12}\tau'_{R,{\rm SHG}}|U_1|^2
  \frac{\partial U_1}{\partial \tau'}\right] 
\\ +\nkerrb^2\left[|U_1|^2U_1-\tau'_{R}U_1
  \frac{\partial |U_1|^2}{\partial \tau'}\right] =0.  
\end{multline}
In Eq.~(\ref{eq:fh-chi2-and-chi3})
self-phase modulation (SPM) effects from both quadratic
and cubic nonlinear effects are present.  For $\Delta k > 0$,
$\chi^{(2)}:\chi^{(2)}$ self-defocusing phase shifts starts to cancel
the $\chi^{(3)}$ self-focusing phase shifts. Thus, for soliton
compression with normal dispersion and self-defocusing phase shifts,
the \textit{effective soliton number}
\begin{align}
\neffs^2&\equiv\nshgb^2-\nkerrb^2
\nonumber\\
\label{eq:Neff}
&=L_{\rm D,1}\Ein^2\frac{\omega_1}{c}\left(\frac{\omega_1}{c\Delta k}
  \frac{\deff^2}{\neff 1 \neff 2} -n_{{\rm Kerr},1} \right),
\end{align}
must be larger than unity, $\neffs>1$. This is analogous to the cubic
soliton compressor, for which compression will only occur for
$\nkerrb>1$. In Sec.~\ref{sec:Numer-results:-BBO} these conclusions
are supported numerically through simulations of
Eqs.~(\ref{eq:shg-bulk-U}). By this we demonstrate that
$\neffs$ governs all the behaviour of the compressed solitons, at
least in the \textit{stationary regime}. This
regime is now discussed.

\subsection{The stationary regime}
\label{sec:stationary-regime-1}

An important observation is that Eq.~(\ref{eq:fh-chi2-and-chi3}) has
a GVM-induced Raman-like term similar to the one
induced by the delayed cubic response in the NLSE~(\ref{eq:nlse-fh}).
Its characteristic dimensionless temporal response is
\cite{ilday:2004,moses:2006}
\begin{align}
  \label{eq:TR-shg}
  \tau'_{R,{\rm SHG}}\equiv
2 |d_{12}|/|\Delta k| \tin,
\end{align}
and $\tau'_{R,{\rm SHG}}=\tau_{R,{\rm SHG}}/\tin$. Typically, 
$\tau_{R,{\rm SHG}}$ lies between 1-5 fs in BBO. This term imposes a Raman-like
red-shift of the FW, which causes the compressed pulse to be
asymmetric and limits the compression achievable.

To understand this better, observe the effective nonlinear FW
phase shift built up during propagation 
$ \phi_{\rm NL}(z)\equiv -{\rm sgn}(\Delta
k)z/\lshgb=-z\omega_1^2\deff^2 \Ein^2/c^2 \neff{1}\neff{2}\Delta k $
\cite{desalvo:1992}, which holds when $\sqrt{|\Delta k|\lshgb}\gg 1$
(weak conversion to the SH). For efficient compression $\phi_{\rm NL}$
should at least be on the order of $-\pi$
\cite{liu:1999,ashihara:2002}, and evidently the pulse needs to
propagate a shorter distance to achieve this for $\Delta k$ low (as
long as the cascading limit is upheld). However, Liu \textit{et al.}
\cite{liu:1999} pointed out that GVM sets a lower limit to $\Delta k$:
In phase-mismatched SHG, the phase between the FW and SH changes
several times, corresponding to a conversion/back-conversion cycle of
energy between FW and SH fields.  The distance over which the relative
phase changes sign once is exactly characterized by the coherence
length $L_{\rm coh}=\pi/|\Delta k|$. Thus, it is important that the
temporal walk-off (GVM) length $\lgvm\equiv \tin/|d_{12}|$ is (much)
larger that the coherence length. Liu \textit{et al.}  \cite{liu:1999}
found that it is sufficient to demand that $L_{\rm coh}<4\lgvm$.
Thereby the so-called \textit{stationary regime} was defined, thus
requiring $\Delta k>4\pi|d_{12}|/\tin$. 
Therefore it is advantageous to have low GVM because it gives access
to large nonlinear phase shifts that occur when $\Delta k$ is not too
large. In addition, low GVM will result in a smaller $\tau'_{R,{\rm
    SHG}}$-term in Eq.~(\ref{eq:fh-chi2-and-chi3}) so the Raman-like
perturbations become smaller. This was a main point of
Ref. \cite{moses:2006}, where it was shown that for a given amount of
GVM the compression increases as the soliton number is increased until
a certain critical value where the Raman-like perturbations start to
dominate. If GVM is reduced the soliton number can be increased further
before this happens, and stronger compression can be achieved. 
We recently devised a more accurate theory for the stationary and
nonstationary regimes \cite{bache:2007a}, showing that the cascaded
nonlinearity gives rise to a nonlocal response, and that in presence
of GVM the nonlocal response function is asymmetric (Raman-like). In
the stationary regime the nonlocal response function is localized and
Eq.~(\ref{eq:fh-chi2-and-chi3}) is recovered in the weakly nonlocal
regime (where the response function is very narrow compared to the
propagating pulse). In the nonstationary regime the nonlocal response
function is oscillatory and does not decay. This has severe
consequences to the built up negative nonlinear phase shift and
results in very poor compression. Finally, a more accurate requirement
for being in the stationary regime
was found as 
\begin{align}
  \label{eq:dkmin-gvm}
  \Delta k>\Delta k_{\rm sr}\equiv d_{12}^2/2\bpp_2, 
\end{align}
which notably is independent of $\tin$. In what follows, we will use
this value. We should note that satisfaction of
inequality~(\ref{eq:dkmin-gvm}) does not eliminate the GVM-induced
perturbation: Raman-like effects
are present when GVM is nonzero both in the stationary regime $\Delta
k>\Delta k_{\rm sr}$, where Eq.~(\ref{eq:fh-chi2-and-chi3}) holds,
and in the nonstationary regime $\Delta k<\Delta k_{\rm
  sr}$. However, in the latter case, where the nonlocal response
function is oscillatory and unbound, the Raman-like perturbations are
stronger.

\subsection{The compression window}
\label{sec:compression-window}

\begin{figure}[tb]
\centerline{
  \includegraphics[width=8.5cm]{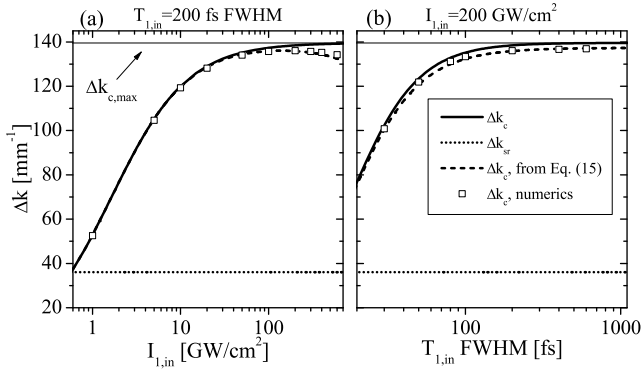}}
\caption{The compression window for BBO with $\lambda_1=1064$~nm using
  (a) $\tin=200$~fs, and (b) $\Iin=200~{\rm GW/cm^2}$. Also shown is
  $\Delta k_c$ found in numerics (see
  Sec.~\ref{sec:Transition-to-compr}), and the dashed line is an
  empirical scaling law~(\ref{eq:energy-fluence-fit}) based on these
  simulations.}
\label{fig:window}
\end{figure}

In the previous sections we have imposed two requirements for clean
soliton compression. First, in order to observe 
solitons, the effective soliton number must
be larger than unity $\neffs>\neffc=1$. This gives from
Eq.~(\ref{eq:Neff}) a critical value of the phase mismatch
\begin{align}
\label{eq:deltakc}
\Delta k<\Delta k_c=\frac{\omega_1\deff^2}{c\neff{1}\neff{2}n_{{\rm
  Kerr},1}(1+\nkerrb^{-2})}.
\end{align}
Second, $\Delta k>\Delta k_{\rm sr}$ to be in the stationary regime,
see Eq.~(\ref{eq:dkmin-gvm}), which is strictly related to the size of
GVM and SH group-velocity dispersion, and is independent of the input
pulse duration. This defines a \textit{compression window}, inside
which we can expect clean compression:
\begin{align}\label{eq:window}
\Delta k_{\rm sr}<\Delta k<\Delta k_c .
\end{align}
Staying inside this window is therefore a question of choosing the
right phase mismatch. Keeping the window open ($\Delta k_{\rm
  sr}<\Delta k_c$) is also a matter of having the right input pulse
intensity and duration, see Fig.~\ref{fig:window}. Moreover, there are
cases where the material nonlinearity balance and/or GVM effects
do not allow for Eq.~(\ref{eq:window}) to be satisfied. For
instance, although $\Delta k_c$ increases with the input intensity,
see Fig.~\ref{fig:window}(a), its maximum value, occurring when
$\nkerrb^{-2}\rightarrow 0$, is $\Delta k_{c,\rm
  max}=\omega_1\deff^2/c\neff{1}\neff{2}n_{{\rm Kerr},1}$. Thus, the
balance between quadratic and cubic nonlinearities is a
fundamental limit on the peak value of $\Delta k_c$. Moreover, when
GVM is large $\Delta k_{\rm sr}$ becomes high, so very
large quadratic nonlinearities are needed to open the window. 
Note that compression can be achieved outside the compression window
when $\Delta k<\Delta k_{\rm sr}$. However, due to the strong
Raman-like effects in the nonstationary regime $\neffs$ must be kept
small in order to get a clean symmetric pulse, allowing only for
moderate compression. Conversely, even inside the compression window
distorted compressed pulses may occur \cite{bache:2007a}.

\subsection{Transverse spatial effects}
\label{sec:Other spatial geometries}

Let us end with a brief discussion of the implications of including
effects of modulation in the transverse spatial dimensions ($x$ and
$y$).  The equations will then include both diffraction terms and
space-time focusing terms (due to steepening effects of diffraction
\cite{brabec:1997}).  In its simplest effect, diffraction will act to
alter the pulse intensity by altering the mode area, which will lead
to changes in the soliton order during propagation.  Therefore, when
$z_0$ is on the order of a characteristic diffraction length or
longer, diffraction will have significant effects on the compression
results.  Additionally, the space-time focusing terms, which couple
temporal and transverse spatial coordinates and become significant at
few-cycle pulse durations, will be expected to produce nontrivial
spatiotemporal distortions.

Moreover, in the plane-wave limit, \textit{i.e.}, where diffraction
effects are small, while the temporal dynamics can be well described
by the model presented in this paper, inhomogeneity in the spatial
intensity profile will result in inhomogeneous compression across the
beam.  For example, for a Gaussian profile only the central part of
the beam may have the appropriate intensity to be efficiently
compressed.  In order to achieve a more homogeneous compression, a
super-Gaussian profile can be used \cite{moses:2007}.

Another case is where the light is guided, such as in an optical
fiber. Here, the nonlinear terms in the propagation equations must
include the effective spatial mode overlap areas, and thus modifying
the soliton numbers as well.  Another important difference in the
guided case is that the dispersion is affected. Moreover, it is
possible to achieve very high intensities with low energy pulses
because the light can be confined in a very small area. We will in
another publication go into details with the specific case of a
photonic crystal fiber, and try to investigate the implications of
wave-guiding effects on the cascaded quadratic soliton compressor.

\section{Numerical results}
\label{sec:Numer-results:-BBO}

We now proceed to the numerical results, where we first study the
transition to compression, and then derive the empirical scaling laws
for the compressed pulses. The simulations were done using
Eqs.~(\ref{eq:shg-bulk-U}), which include self-steepening, the SEWA
corrected SH dispersion term and higher-order nonlinear mixing terms.
The coupled equations were solved using a split-step Fourier technique
with a 2nd order Runge-Kutta algorithm to evaluate the nonlinear terms
in the time domain. The dispersion terms were evaluated in the
frequency domain, and since $k_j(\omega)$ is known analytically (from
the Sellmeier equations of BBO \cite{dmitriev:1999}) we can actually
evaluate the dispersion operator $\hat D_j$ exactly \cite{dudley:2006}
instead of using the expansions~(\ref{eq:Dprop}). For the effective
SH operator~(\ref{eq:D2-eff-dimless-final-app}) we used 30 terms in
the expansion of $\hat S_2^{-1}$.
The steepening terms were applied by using the convolution theorem and
thus letting the steepening operator act on the nonlinear terms in the
frequency domain.  The number of discretization points $N_z$ in the
$z$-direction was chosen so that around 15 steps were taken within a
single coherence length $L_{\rm coh}=\pi/|\Delta k|$.  The number of
temporal points were $2^{12}$-$2^{15}$, primarily dictated by the fact
that due to GVM the SH will have a trailing pulse, which must stay
inside the time window of the simulations.

We decided to use BBO as the nonlinear medium  for
reasons described in App.~\ref{sec:BBO}. We chose the
Yb/Nd:YAG wavelength $\lambda_1=1064~$nm.
At this wavelength BBO has a medium level of GVM ($d_{12}\simeq
-100$~fs/mm). If instead we consider $\lambda_1=800~$nm -- another
wavelength of considerable interest -- GVM is roughly twice as big and
the compression window is much smaller.  From this point of view
$\lambda_1=1064~$nm is better suited for this investigation.

\begin{figure}[tb]
\centerline{\includegraphics[width=8cm]{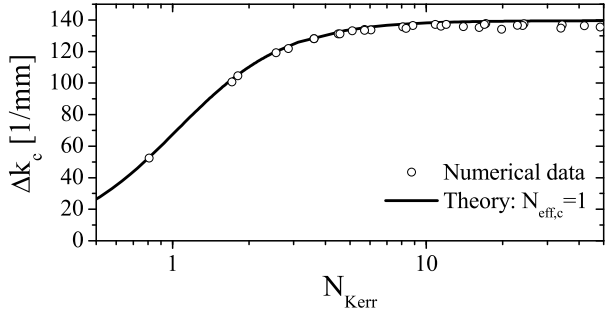}}
\caption{Locating the critical transition point to pulse
  compression. The critical phase mismatch $\Delta k_c$ vs.  $\nkerrb$
  in a semi-log plot for simulations having $\Iin=1-600~{\rm GW/cm^2}$
  and $\tin=80-600$~fs FWHM.}
\label{fig:neff}
\end{figure}

\subsection{Transition to compression}
\label{sec:Transition-to-compr}

First, we characterize the system at the transition to
compression, in order to understand whether the requirement
$\neffs>\neffc=1$ holds. Quantitatively, we first found the critical
phase mismatch $\Delta k_c$ for a given $\Iin$ and $\tin$ (thus,
$\nkerrb$ is fixed, while $\nshgb$ is varied as $\Delta k$ is
scanned). This was done through numerics of $L=50$ mm propagation in a
BBO crystal.  First, we located a $\Delta k$ just before the
transition (slight decrease in pulse intensity due to pulse
broadening) and a $\Delta k$ just after the transition (slight
increase in pulse intensity due to pulse narrowing). $\Delta k_c$ was
then found by interpolation. 
We chose sech-shaped input pulses $U_{1,\rm in}={\rm sech}(\tau')$
because they are solitonic solutions to the NLSE (without higher order
dispersion and nonlinearities), and the pulse intensities during
propagation were correspondingly fitted to a sech$^2$-shaped pulse.
Based on Ref. \cite{chen:2002} we expect the results presented here to
remain largely identical for sech- and Gaussian-shaped input pulses.

\begin{figure}[tb]
\centerline{\includegraphics[width=8cm]{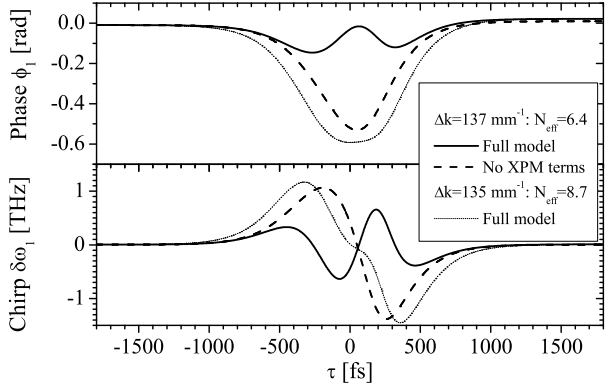}}
\caption{Numerical simulations showing the FW phase and chirp after
  propagation in 50 mm BBO with a high fluence $\Phi_{1,\rm
    in}=272~{\rm mJ/cm^2}$ resulting in $\neffc=8.4$. Input:
  $\Iin=400~{\rm GW/cm^2}$ and $\tin=600$~fs FWHM. }
\label{fig:below-neffc}
\end{figure}

In Fig.~\ref{fig:neff}(a) we plot $\Delta k_c$ vs. $\nkerrb$ for many
different values of $\tin$ and $\Iin$. As
Sec.~\ref{sec:soliton-numbers} suggested, the transition to
compression can expressed as $\neffs>\neffc=1$, and the corresponding
theoretical line calculated from Eq.~(\ref{eq:deltakc}) is also shown.
We observe that the prediction $\neffc=1$ only holds for low values
$\nkerrb$, while for larger values the simple requirement $\neffs>1$
is not enough. A careful study revealed that for large values of
$\tin$ and $\Iin$ the critical effective soliton number scales as
$\neffc\propto \tin\Iin$.
\begin{figure}[tb]
\centerline{\includegraphics[width=8cm]{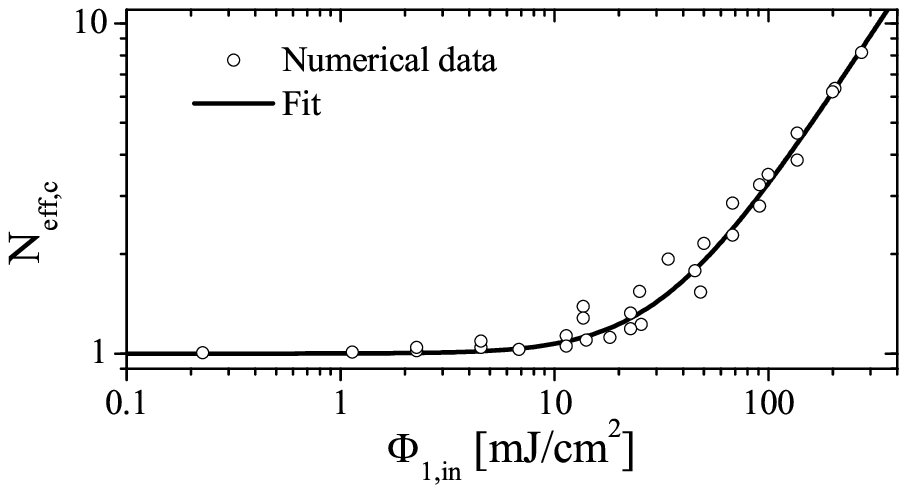}}
\caption{The data from Fig.~\ref{fig:neff} shown vs. the input energy
  fluence $\Phi_{1,\rm in}=2\tin\Iin$ in a log-log plot.}
\label{fig:neff-uin}
\end{figure}
This suggests that the effect is not related to any of the soliton
numbers (which scale as $\tin\sqrt{\Iin}$). Instead, consider the
energy fluence $\Phi_j=\int dt I_j(t)$, which for a sech-shaped pulse
$I_0{\rm sech}(t/T_0)$ is $\Phi=2T_0 I_0$. In Fig.~\ref{fig:neff-uin}
the corresponding $\neffc$ 
from the $\Delta k_c$ data of Fig.~\ref{fig:neff} is plotted vs.
$\Phi_{1,\rm in}$, and the data are well represented by the rational fit
\begin{align}
  \label{eq:energy-fluence-fit}
\neffc=1+\frac{\Phi_{1,\rm in}'}{1+1/\Phi_{1,\rm in}'}, \quad
\Phi_{1,\rm in}'=\Phi_{1,\rm in}/\Phi_c,
\end{align}
where the free fitting parameter $\Phi_c=33.0~{\rm mJ/cm^2}$. We
expect that this $\Phi_c$ value is only valid for the particular case
studied. 
A possible explanation for the deviation from $\neffc=1$ was found by
studying the initial stages of propagation when the fluence is high
such that $1\ll \neffs< \neffc$, see Fig.~\ref{fig:below-neffc} for
$\Delta k=137~{\rm mm^{-1}}$ and $\Phi_{1,\rm in}=272~{\rm mJ/cm^2}$.
As one would expect when $\neffs>1$ a substantial negative FW phase
shift builds up. However, in the pulse center an increase in the phase
is seen, which causes the chirp across the pulse center to be
non-monotonic making the pulse unable to compress upon further
propagation. By turning off the Kerr cross-phase modulation (XPM)
terms in the numerics, we observe that this phase increase in the
pulse center disappears; now the pulse has a monotonic chirp across
the pulse center so it can compress. This was indeed observed upon
further propagation.  Turning on the Kerr XPM terms again and taking
$\Delta k=135~{\rm mm^{-1}}$ to achieve $\neffs> \neffc\gg 1$,
$\neffs$ is now large enough cancel also the detrimental contribution
from the Kerr XPM terms: the phase increase in the center is no longer
there making the chirp monotonic. Therefore the pulse can compress,
which we indeed observed. Thus, the deviation from $\neffc=1$ comes
from a positive phase contribution in the pulse center, which seems to
originate from the Kerr XPM terms. These cannot be neglected due to
the high fluence, but were indeed neglected in the NLSE-type
model~(\ref{eq:fh-chi2-and-chi3}) from which the conjecture $\neffc=1$
originated.  An indication of this high-fluence effect on the XPM
terms was recently observed experimentally \cite{moses:2007a}.
Finally, as the Kerr XPM coefficient is changed we found that
$\Phi_c\propto B^{-1/2}$.

We can conclude that $\neffc$ deviates quite strongly from unity when
the fluence (and thus the soliton numbers) is large. However, 
for $\Phi_{1,\rm in}\lesssim \Phi_c$, $\neffc=1$ can be used as a safe
estimate of the transition point.


\subsection{Scaling laws for compression parameters }
\label{sec:Scal-laws-compr}

We proceed to investigate how the soliton compressor behaves when
compression is successful. We seek to find general scaling laws that
can tell us at what position $\lmax$ in the crystal the pulse reaches
its optimal single-spike compressed state -- \textit{i.e.}, with
maximum intensity and minimum duration -- and to check its duration,
compression factor, and quality.

For soliton compression in the NLSE Mollenauer \textit{et al.}
\cite{mollenauer:1983} first studied the compression for
$1<\nkerrb<15$.  Based on this Dianov \textit{et al.}
\cite{dianov:1986} showed the following empirical scaling law for the
optimal compression length
\begin{align}
\label{eq:lmaxz0-dianov}  
\frac{\lmax}{z_0}=\frac{0.32}{\nkerrb}+\frac{1.1}{\nkerrb^2}, \quad
  10<\nkerrb < 50,
\end{align}
where $z_0=\tfrac{\pi}{2} L_{\rm D,1}$ is the soliton
length \cite{agrawal:1989}. Also based on Ref.~\cite{mollenauer:1983},
Tomlinson \textit{et al.} \cite{tomlinson:1984} reported the empirical
scaling law for the compression factor
$f_c\equiv\tin/T_{1,\rm opt}$
\begin{align}
\label{eq:fc-dianov}  
  f_c&=4.1\nkerrb, \quad 1\ll\nkerrb<50.
\end{align}
The validity range was reported in Ref.~\cite{dianov:1986}.
Finally, the pulse quality is defined as the fluence in a sech$^2$-fit
to the central spike relative to the input fluence $Q_c\equiv \Phi_{1,\rm
  sech-fit}/\Phi_{1,\rm in} 
$, which is equivalent to taking $Q_c=I_1/\Iin f_c$. Thus, we should
expect that $Q_c\propto 1/f_c$ \cite{agrawal:2001}.  As we will now 
show, the cascaded quadratic soliton compressor follows nicely
these scaling laws when $\nkerrb$ is replaced by $\neffs$.
This holds as long as we are in
the stationary regime, \textit{i.e.}, when the NLSE-like model
Eq.~(\ref{eq:fh-chi2-and-chi3}) is a good approximation to the system
and when GVM effects are small.

\begin{figure}[t]
\centerline{\includegraphics[height=4.2cm]{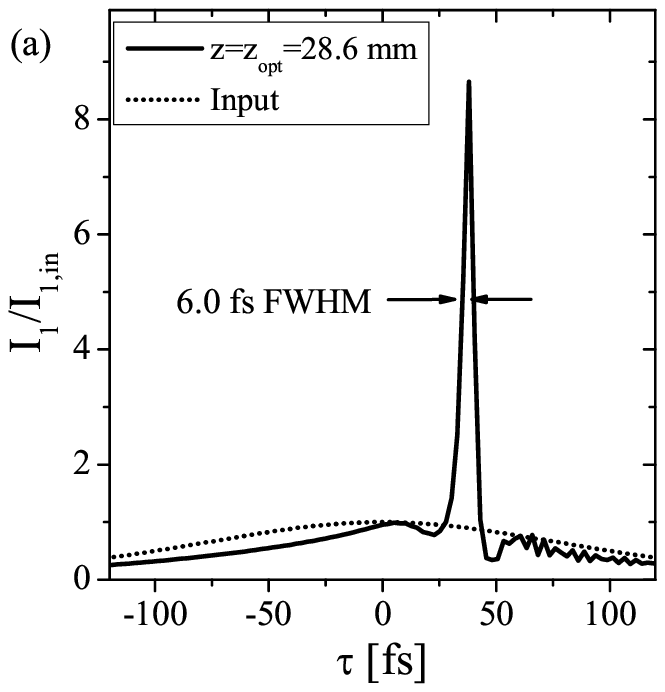}
  \includegraphics[height=4.2cm]{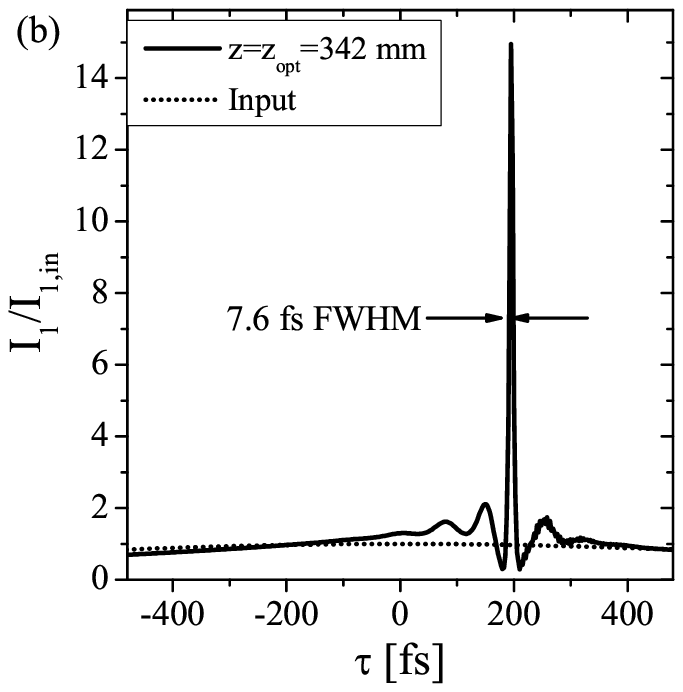}}   
\caption{Selected simulations of clean compressed pulses at the
  optimal compression length. (a) $\Delta k=50~{\rm mm^{-1}}$,
  $\tin=200$~fs FWHM and $\Iin=59~{\rm GW/cm^2}$ ($\neffs=8$)
  resulting in a 6.0 fs pulse ($f_c=33$) with $Q_c=0.26$. (b) $\Delta
  k=55~{\rm mm^{-1}}$, $\tin=2000$~fs FWHM and $\Iin=26.7~{\rm
    GW/cm^2}$ ($\neffs=50$) resulting in a 7.6 fs pulse ($f_c=264$)
  with $Q_c=0.06$.}
\label{fig:compression-examples}
\end{figure}

We performed a wide range of simulations using sech-shaped un-chirped
input pulses with parameters ranging from $\tin=50-2000$~fs and
$\Iin=10-400~{\rm GW/cm^2}$. The simulations used in the following
resulted in cleanly compressed pulses as demonstrated in
Fig.~\ref{fig:compression-examples}. It is worth to stress that the
depletion of the FW due to conversion into SH was quite low in all
cases ($<3\%$ in the stationary regime) due to the large values of
phase mismatch $40<\Delta k<120$ accessible in the compression window.
Some selected compression examples are given in
Fig.~\ref{fig:compression-examples}: (a) shows a typical case with a
200 fs input, where we have optimized the phase mismatch and the input
intensity to give a 6.0 fs compressed pulse.  This implies a
compression ratio of $f_c=33$ and the pulse quality is $Q_c=0.26$.
The remaining pulse energy resides in the unwanted pedestal as well as
in the SH (2.3\% conversion occurred). As a more extreme case, (b)
shows a 2 ps long pulse compressed to a clean 8.3 fs pulse, implying
an impressive $f_c=264$.  A lot of the energy remains in the pedestal
(only 0.5\% is converted to the SH), and the pulse quality is merely
$Q_c=0.06$.  This is typical of large compression ratios. Note the
difference in $\lmax$; $\lmax=342~$mm in
Fig.~\ref{fig:compression-examples}(b) is long, but may be realized
using multiple crystals.

\begin{figure}[t]
\centerline{\includegraphics[width=8.5cm]{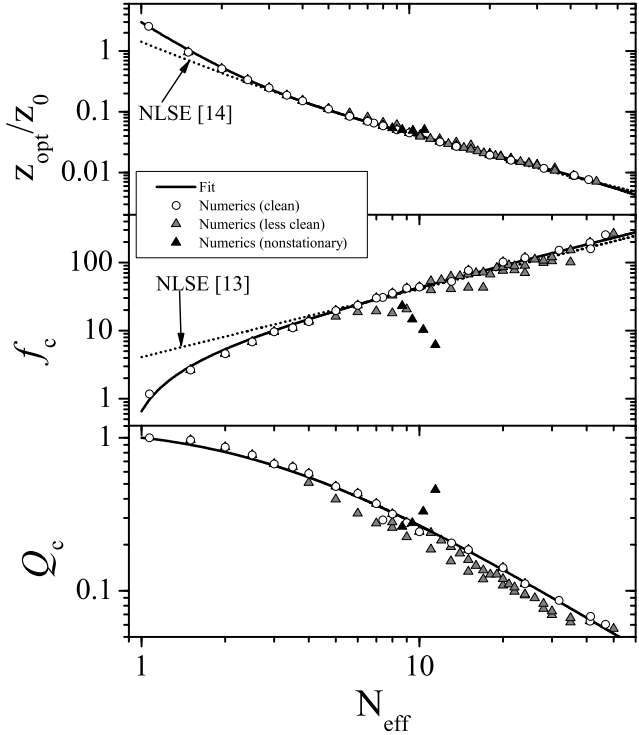}}
\caption{Results of numerical simulations showing the optimum
  compression parameters vs. $\neffs$ in log-log plots for (a) the
  compression length $\lmax/z_0$, (b) the compression factor $f_c$ and
  (c) the compression quality $Q_c$. The simulations marked with round
  symbols resulted in a clean compressed pulse, while the triangles
  resulted in less clean pulses. The four black triangles gradually
  enter the
  nonstationary regime. 
The solid lines are fits to the clean data
  [Eqs.~(\ref{eq:zopt-fit})-(\ref{eq:qc-fit})], while the dotted
  lines are the scaling laws~(\ref{eq:lmaxz0-dianov})
  and~(\ref{eq:fc-dianov}). }
\label{fig:lmax}
\end{figure}

In Fig.~\ref{fig:lmax}(a) we show $\lmax/z_0$ as function of the
effective soliton number $\neffs$. The data follow the scaling
law~(\ref{eq:lmaxz0-dianov}) -- with $\nkerrb$ is replaced by
 $\neffs$ -- quite well, but there are deviations for
small $\neffs$. This is due to the $N^{-2}$ term in
Eq.~(\ref{eq:lmaxz0-dianov}), so an improved fit gave 
\begin{align}
  \label{eq:zopt-fit}
\frac{\lmax}{z_0}=\frac{0.44}{\neffs}+\frac{2.56}{\neffs^3}-0.002.
\end{align}
We have also included data points where the compressed
pulses were less clean (either having trailing or leading
oscillations, or being somewhat asymmetric). As the plot indicates
they start kicking in when $\neffs>10$, which can be explained by 
increased XPM contributions as well as increased influence of the 
GVM induced Raman-like perturbation.  Nonetheless,
these less clean data still follow the scaling law~(\ref{eq:zopt-fit})
closely. An exception is when the nonstationary regime is close: We
included four data points (black symbols), which have the same
parameters as Fig.~\ref{fig:compression-examples}(a) and where $\Delta
k$ is gradually decreased so the final data point is in the
nonstationary regime. These data
points start to deviate from the scaling laws because the compressed
pulses experience more strongly the GVM induced Raman-like effects. 

In Fig.~\ref{fig:lmax}(b) we plot the compression factor $f_c$ as
function of $\neffs$. Again the scaling law~(\ref{eq:fc-dianov}) holds
well except for small $\neffs$. A linear fit to the clean data gave
the following scaling law valid also for small $\neffs$
\begin{align}
  \label{eq:fc-fit}
f_c=4.7(\neffs-0.86).
\end{align}
The less clean data also follow the scaling law~(\ref{eq:fc-fit})
quite well, but the data points approaching the nonstationary regime
(black triangles) very quickly separate out. Thus, $f_c$ is very
sensitive to the effects of the GVM induced Raman-like effects in the
nonstationary regime. 

As the last dimensionless parameter, we show in Fig.~\ref{fig:lmax}(c)
the pulse compression quality $Q_c$ as a function of $\neffs$. The
data roughly follow a rational function 
\begin{align}
  \label{eq:qc-fit}
Q_c=[0.24 (\neffs-1)^{1.11}+1]^{-1}.
\end{align}
The exponent 1.11 deviates from unity due to the behaviour of $Q_c$ for
small $\neffs$. However, for large $\neffs$ we find $Q_c\propto
f_c^{-1}$, as predicted. 

\begin{figure}[tb]
\centerline{\includegraphics[width=8.5cm]{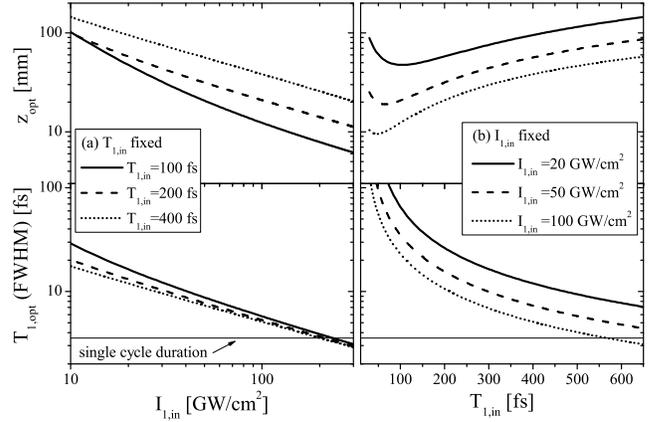}}
\caption{The optimal compressor length and the expected compressed
  pulse duration in a BBO for $\lambda_1=1064$~nm and $\Delta k=50 ~{\rm
  mm^{-1}}$ with (a) $\tin$ fixed; (b) $\Iin$ fixed.}
\label{fig:physical-scaling}
\end{figure}

As a more concrete example Fig.~\ref{fig:physical-scaling} shows the
compressor length one should choose for optimal compression $\lmax$,
together with the expected compressed pulse duration. These were
calculated from Eqs.~(\ref{eq:zopt-fit}) and (\ref{eq:fc-fit}) for
$\Delta k=50 ~{\rm mm^{-1}}$. Single-cycle pulses are available for
$\Iin>200~{\rm GW/cm^2}$ with realistic compressor lengths around
$5-20$~mm.

\begin{figure}[tb]
\centerline{\includegraphics[width=7cm]{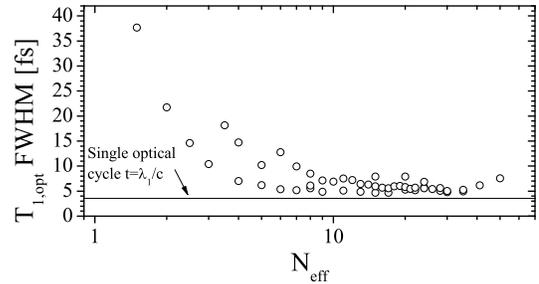}}
\caption{The duration of the compressed pulse vs. $\neffs$ in a log
  plot. The data are from the same simulations as in
  Fig.~\ref{fig:lmax}, but only those with $\Delta k=55 ~{\rm mm^{-1}}$
  are shown. }
\label{fig:tout-neff}
\end{figure}

Finally, Fig.~\ref{fig:tout-neff} summarizes output pulse duration vs.
$\neffs$, providing a sense of the possible pulse compression in BBO
at $\lambda_1=1064$ nm. The lowest value observed was 4.7 fs
FWHM, corresponding to 1.3 optical cycles.  Single-cycle pulses were
not observed, most likely due to GVM-induced Raman
effects and Kerr XPM effects.

\section{Conclusions}
\label{sec:Conclusions}

In summary, we find that the effective soliton number
$\neffs=\sqrt{\nshgb^2-\nkerrb^2}$ is the proper dimensionless
quantity for describing soliton compression using cascaded quadratic
nonlinearities in the stationary regime. Soliton compression generally
only occurs when $\neffs>1$, and this in turn requires that the
quadratic nonlinearity dominates over the cubic nonlinearity. In this
balance, material and input pulse parameters such as the phase
mismatch, the input pulse intensity and duration play a crucial role.
When the pulse energy is large $\neffs>1$ is no
longer a sufficient demand. We attributed this to Kerr XPM effects,
and we found an empirical scaling law relating $\neffc$ to the input
fluence.

We showed through a large number of realistic numerical simulations of
a BBO crystal that the system clearly obeys dimensionless scaling laws
dictating the optimal compression propagation distance, the pulse
compression factor and quality. These scaling laws were expressed
through the effective soliton number $\neffs$, and are very similar to
the ones observed in the NLSE. The scaling laws are general and hold
also for other materials and wavelengths, and will serve as a crucial
tool to determine input pulse peak intensity and duration as well as
phase-matching conditions and optimal crystal lengths.  

Besides the requirement of a strong quadratic nonlinearity, avoiding
the nonstationary regime, where GVM-effects are strong, is another
obstacle to observe the desired compression in the considered system.
While compression may still occur, the final pulse is too distorted to
be of any use due to GVM induced Raman-like effects. This is
particularly a problem for short pulses, high intensity pulses, and in
materials with a weak quadratic nonlinearity and/or large GVM. In this
paper we considered the stationary regime, where this effect is
negligible and stress that the scaling laws for compression hold only
in this regime. In subsequent publications we will study the GVM
effects in more detail, and show that the cascaded quadratic
nonlinearity induces a nonlocal response on the FW, and that in
presence of GVM this response becomes Raman-like \cite{bache:2007a}.
Moreover, in the nonstationary regime the response function is no
longer localized, which has severe implications on the compressed
pulse.

Let us finally touch on the exciting prospect of the cascaded
quadratic soliton compressor in a wave-guiding configuration. As
mentioned, GVM effects are detrimental to the compression performance,
and these become even more pronounced for shorter input wavelengths
such as 800 nm because GVM is much stronger. However, in a photonic
crystal fiber (PCF) wave-guiding effects can dramatically alter the
dispersion, and in particular it was shown that for SHG it is possible
to achieve zero GVM \cite{bache:2005a}.  Moreover, in a PCF the mode
areas are very small (roughly a few $\mu{\rm m^2}$ is possible) so
very high peak intensities are possible even with very low pulse
energies. This will allow achievement of low-energy clean compressed
pulses, once issues with quadratic nonlinear response of optical
fibers have been resolved.

\section{Acknowledgments}
\label{sec:Acknowledgments}

M.B. was supported by The Danish Natural Science Research Council
Grant 21-04-0506. J.M. and F.W.W. were supported by NSF Grants
PHY-0099564 and ECS-0217958. O. Bang is acknowledged for discussions.

\appendix
\section{The cubic nonlinear response}
\label{sec:cubic-nonl-resp}

We here show how a cubic nonlinearity is included
in the model. 
The cubic nonlinear polarization response is
\begin{multline}
  \label{eq:pol}
  \Pol_{\rm NL}^{(3)}=\evac\iiint_{-\infty}^\infty
  dt_1dt_2dt_3 
\\\times
\underline{\underline{\chi}}^{(3)}(t-t_1,t-t_2,t-t_3)\vdots
  \E(t_1) \E(t_2) \E(t_3) ,
\end{multline}
$\underline{\underline{\chi}}^{(3)}$ is a rank 4 tensor describing the
cubic nonlinear response of the material, which we let take the
functional form $ \underline{\underline{\chi}}^{(3)}(t_1,t_2,t_3)=
\underline{\underline{\chi}}^{(3)}R(t_1)\delta(t_2)\delta(t_3) $,
where $R(t)$ is the normalized material Kerr response function.  In
arriving at Eqs.~(\ref{eq:shg-bulk-E}) we then assumed the waves
monochromatic and polarized along arbitrary directions $\E(t)={\rm
  Re}[{\mathbf x}_1 E_1(t)e^{-i\omega_1 t}+ {\mathbf
  x}_2E_2(t)e^{-i\omega_2 t} ]$, 
where ${\mathbf x}_j$ is the unit polarization vector. In SHG the FW
and SH waves are either all polarized along the same direction (Type
0), or the two FW are polarized along the same direction but
orthogonal to the SH (Type I), or finally the two FW photons could be
polarized orthogonally to each other while the SH is parallel with one
of the FW photons (Type II) \cite{dmitriev:1999}. The propagation
equations presented here are valid for Type 0 and I, since the FW
photons are considered degenerate. We assume an isotropic cubic Kerr
nonlinearity, for which the only nonzero tensor
components of $\underline{\underline{\chi}}^{(3)}$ are
$\chi^{(3)}_{xxyy}=\chi^{(3)}_{xyyx}=\chi^{(3)}_{xyxy}=\chi^{(3)}_{xxxx}/3$
\cite{agrawal:2001}. This means that the XPM coefficient in
Eq.~(\ref{eq:shg-bulk-fh-raman}) is $B=2$ for Type 0 SHG, and $B=2/3$
for Type I SHG. We now divide the material Kerr response into an
electronic response and a vibrational Raman response
$R(t)=(1-f_R)\delta(t)+f_Rh_R(t)$ \cite{agrawal:1989},
where $f_R$ is the fractional contribution of the vibrational Raman
response. The first part describes the electronic response, which can
be considered instantaneous, resulting in the cubic SPM and XPM terms
in
Eqs.~(\ref{eq:shg-bulk-E}). 
The vibrational Raman response for a general 3-wave mixing process
with two different frequencies is \cite{kumar:1993}
\begin{multline}
\label{eq:Raman}
  \Rm_j(\tau)\equiv \hat S_j\int_{-\infty}^\infty ds
  h_R(s)
\\ \times\Big\{E_j(\tau) \left[
  |E_j(\tau-s)|^2+\tfrac{1}{2}B|E_m(\tau-s)|^2\right] 
\\ + \tfrac{1}{2}B E_j(\tau-s)E_m^*(\tau-s)e^{i(\omega_j-\omega_m)s}E_m(\tau)
\Big\},
\end{multline}
where $j=1,2$ and $m=3-j$, and only terms that are phase matched are
included \cite{kumar:1993}.  This 
Raman response is governed by the gain function $h_R$. For SHG, the
integral over the term containing $e^{i(\omega_j-\omega_m)s}$ is
vanishing since it implies evaluation of $h_R$ is evaluated at frequency
offset $|\omega_j-\omega_m|=\omega_1$ much larger than the
typical spectral width of $h_R$ (in the THz range). Therefore,
for SHG only the term $E_j(\tau) \left[
  |E_j(\tau-s)|^2+\tfrac{1}{2}B|E_m(\tau-s)|^2\right] $ remains, 
and the generalized coupled
equations~(\ref{eq:shg-bulk-E}) become 
\begin{subequations}
\label{eq:shg-bulk-E-explicit}
\begin{align}
  \label{eq:shg-bulk-fh-raman-explicit}
  &\Lm_1E_1+\kappa_{\rm SHG,1}^E\hat S_1E_1^*E_2e^{i\Delta k
    z}
 \\\nonumber  &+\kappa_{\rm Kerr,1}^E \hat S_1E_1\Bigg\{(1-f_R)
  \left(|E_1|^2+B|E_2|^2\right) 
\\\nonumber  &+f_R\int
_{-\infty}^\infty ds
  h_R(s)\left[ |E_1(\tau-s)|^2+\tfrac{1}{2}B|E_2(\tau-s)|^2\right]\Bigg\}=0,
\nonumber\\
  \label{eq:shg-bulk-sh-raman-explicit}
  &\Lm_2E_2+\kappa_{\rm SHG,2}^E\hat S_2
E_1^2e^{-i\Delta k
    z}
 \\\nonumber  &+\kappa_{\rm Kerr,2}^E \hat S_2
E_2\Bigg\{(1-f_R)
  \left(|E_2|^2+B|E_1|^2\right) 
\\\nonumber  &+f_R\int
_{-\infty}^\infty ds
  h_R(s)\left[ |E_1(\tau-s)|^2+\tfrac{1}{2}B|E_2(\tau-s)|^2\right]\Bigg\}=0.
\nonumber
\end{align}
\end{subequations}

Finally, the Raman overlap integral has an approximate
form for short, but not extremely
short, pulses
\begin{multline}
  \label{eq:Raman-approx}
  f_R\Rm_j(\tau)\simeq 
f_R\hat S_j E_j(\tau)\left[ |E_j(\tau)|^2+\tfrac{1}{2}B|E_m(\tau)|^2\right]\\
-\tau_R\frac{\partial}{\partial \tau}\left[
  |E_j(\tau)|^2+\tfrac{1}{2}B|E_m(\tau)|^2\right] ,
\end{multline}
where we have used that $\int_{-\infty}^\infty dt h_R(t)=1$, and defined the
well-known Raman response time $\tau_R\equiv f_R\int_{-\infty}^\infty
dt\, th_R(t)$.
Eq.~(\ref{eq:Raman-approx}) is related to intra- and inter-pulse Raman
scattering (IIRS), and
Eqs.~(\ref{eq:shg-bulk-E-explicit}) become
\begin{subequations}\label{eq:shg-bulk-Raman-app}
\begin{align}
  \label{eq:shg-bulk-Raman-app-fh}
  \Lm_1E_1&+  \kappa_{{\rm SHG},1}^E\hat S_1 E_1^*E_2e^{i\Delta k
  z}    
\\\nonumber&+
\kappa_{{\rm Kerr},1}^E\Big\{\hat S_1
  E_1\left[|E_1|^2+B(1-\tfrac{1}{2}f_R)|E_2|^2\right]  
\\\nonumber&\phantom{+\kappa_{{\rm Kerr},1}^E\Big\{}  
-\tau_RE_1\frac{\partial} {\partial
  \tau}\left(|E_1|^2+\tfrac{1}{2}B|E_2|^2\right) 
  \Big\}=0,\\ 
  \label{eq:shg-bulk-Raman-app-sh}
  \Lm_2E_2&+  \kappa_{{\rm SHG},2}^E\hat S_2E_1^2e^{-i\Delta k z}
\\\nonumber&  +\kappa_{{\rm Kerr},2}^E\Big\{\hat S_2
  E_2\left[|E_2|^2+B(1-\tfrac{1}{2}f_R)|E_1|^2\right]  
\\\nonumber&\phantom{ +\kappa_{{\rm Kerr},2}^E\Big\{}  
-\tau_RE_2\frac{\partial} {\partial
  \tau}\left(|E_1|^2+\tfrac{1}{2}B|E_2|^2\right) 
  \Big\}=0.
\end{align}
\end{subequations}
Here we have neglected the 2nd order derivatives from
applying self-steepening to the IIRS
term in Eq.~(\ref{eq:Raman-approx}).

\section{Extending the propagation equations to the SEWA regime}
\label{sec:Appendix}

Here we discuss the difference between the slowly varying envelope
approximation (SVEA) and the SEWA propagation equations. SEWA
\cite{brabec:1997} is a general spatio-temporal model that describes
spatio-temporal pulse propagation down to single-cycle pulse
durations. It was recently derived for SHG by Moses and Wise
\cite{moses:2006b}. SEWA does not pose any direct constriction on the
pulse bandwidth, whereas the extended SVEA (with steepening terms and
the general Raman convolution response) holds for \cite{blow:1989}
$\Delta \omega/\omega<1/3$.  Neglecting transverse spatial terms, the
only term the SHG extended SVEA model does not include is related to
the dispersion of the SH, since due to GVM the SH dispersion
operator~(\ref{eq:Dprop}) must be replaced by the following
\textit{effective} operator \cite{moses:2006b} $ \Dm_{2,{\rm
    eff}}\equiv\Dm_2+\hat S_2^{-1}\frac{d_{12}^2}{2k_2}
\frac{\partial^2 }{\partial\tau^2}$. Imposing the scalings
$z'=z/\ld$ and $\tau'=\tau/\tin$, the dimensionless operator
becomes
\begin{align}
  \label{eq:D2-eff-dimless}
  \Dm'_{2,{\rm eff}}&\equiv\Dm_2'+\hat
  S_2'^{-1}\frac{\nu}{2}\frac{\partial^2 }{\partial\tau'^2} ,
\end{align}
where the dimensionless factor $
\nu\equiv cd_{12}^2/\omega_2n_2
  |\bpp_1|$.
Expanding the inverse steepening operator $\hat
  S_2'^{-1}=\sum_{m=0}^\infty
  \left(\frac{-is'}{2}\right)^m\frac{\partial ^m}{\partial \tau'^m},$
we get
\begin{align}
  \label{eq:D2-eff-dimless-final-app}
  \Dm'_{2,{\rm eff}}=\sum_{m=2}^\infty i^m\left[
  \delta_2^{(m)}+\frac{\nu}{2}\left(\frac{s'}{2}\right)^{m-2}\right]
\frac{\partial ^m}{\partial \tau'^m}.
\end{align}

In the SEWA there is no restriction on the pulse bandwidth so
single-cycle temporal resolution is achieved.  However, since we
consider SHG we should be careful. One assumption made when deriving
Eqs.~(\ref{eq:shg-bulk-E}) is namely that the spectra of the FW and SH
do not overlap (substantially). This assumption allows us to separate
the fields as shown in App.~\ref{sec:cubic-nonl-resp}. We chose
$\Delta \omega/\omega_j=0.7$. This gives some overlap between the FW
and SH spectra, but we always made sure that the spectral components
in the overlapping regions were negligible.

\section{BBO}
\label{sec:BBO}

We use a $\beta$-${\rm BaB_2O_4}$ (beta-barium-borate,
BBO) nonlinear crystal, where collinear Type I SHG is possible through
the interaction ${\rm oo\rightarrow e}$, \textit{i.e.} the FW photons
are ordinarily polarized, while the generated SH photon is
extraordinarily polarized.  BBO is a uni-axial crystal where phase
matching can be achieved by birefringent phase matching by changing
the angle $\theta$ between the FW input and the optical $z$-axis of
the crystal. $\deff$ also changes with $\theta$, and is on average
2.22 pm/V in the area in which we are interested in. The dispersion is
calculated from the Sellmeier equations \cite{dmitriev:1999}. Note in
this connection that the analytical transition to compression
$\neffc$, as calculated from 
Eq.~(\ref{eq:deltakc}), is on implicit form.

BBO is an excellent nonlinear medium for the current purpose because
it has a decent quadratic nonlinear strength, and perhaps more
important, it has a small GVM at NIR wavelengths, which is the major
reason for using BBO instead of, \textit{e.g.}, periodically poled
LiNbO$_3$ for these wavelengths.  BBO also has a very low cubic
nonlinear refractive index.  As we noted from Eq.~(\ref{eq:Neff}) the
balance between these is crucial.  
BBO also has a very low two-photon absorption
coefficient (except in the ultra-violet part of the spectrum
\cite{desalvo:1996}), justifying the approximation made in the
derivation of Eqs.~(\ref{eq:shg-bulk-U}).

In the literature several values for the nonlinear refractive index
have been reported. We chose to use $n_{{\rm Kerr},1}^I= 3.65\pm 0.6
\cdot 10^{-20}~{\rm m^2/W}$ reported in
Ref.~\onlinecite{sheik-bahae:1997} mainly because the measurements are
done at phase-matching angles close to ours and also used fs pulses,
but also because this value in the past has given the best agreement
with experimental observations.



\end{document}